\documentclass[twocolumn,twoside,superscriptaddress,pra,aps]{revtex4}
\usepackage{amsmath,mathrsfs,amsbsy,amssymb,lmodern,graphicx,bm,amsthm,amsfonts}
\usepackage{units}
\usepackage{bbm}
\usepackage{color}
\usepackage{txfonts}

\newcommand{\idol}{\ensuremath{\mathbbm 1}}

\newcommand{\tr}{{\rm Tr}}

\begin{document}

\title{Evaluation of entanglement measures by a single observable}
\author{Chengjie Zhang}
\email{zhangchengjie@suda.edu.cn}
\affiliation{College of Physics, Optoelectronics and Energy, Soochow University, Suzhou, 215006, China}
\affiliation{Centre for Quantum Technologies, National University of Singapore, 3 Science Drive 2, Singapore 117543, Singapore}
\author{Sixia Yu}
\affiliation{Centre for Quantum Technologies, National University of Singapore, 3 Science Drive 2, Singapore 117543, Singapore}
\affiliation{Hefei National Laboratory for Physical Sciences at Microscale and Department of Modern Physics,  University of Science and Technology of China, Hefei, Anhui 230026, China}
\author{Qing Chen}
\affiliation{Centre for Quantum Technologies, National University of Singapore, 3 Science Drive 2, Singapore 117543, Singapore}
\affiliation{Department of Physics, Yunnan University, Kunming, 650091, China}
\author{Haidong Yuan}
\affiliation{Department of Mechanical and Automation Engineering, The Chinese University of Hong Kong, Hong Kong}
\author{C.H. Oh}
\affiliation{Centre for Quantum Technologies, National University of Singapore, 3 Science Drive 2, Singapore 117543, Singapore}
\affiliation{Physics Department, National University of Singapore, 3 Science Drive 2, Singapore 117543, Singapore}

\begin{abstract}
We present observable lower bounds for several bipartite entanglement measures including entanglement of formation, geometric measure of entanglement, concurrence, convex-roof extended negativity, and G-concurrence. The lower bounds facilitate estimates of these entanglement measures for arbitrary finite-dimensional bipartite states. Moreover, these lower bounds can be calculated analytically from the expectation value of a single observable. Based on our results, we use several real experimental measurement data to get lower bounds of entanglement measures for these experimentally realized states. In addition, we also study the relations between entanglement measures.
\end{abstract}
\date{\today}

\pacs{03.67.-a, 03.65.Ta, 03.67.Lx}

\maketitle

\section{Introduction}
Quantum entanglement is widely recognized as a valuable resource in quantum information processing. However, it is far from simple to fully determine entanglement. Therefore, the characterization and quantification of entanglement become fundamental problems in quantum information theory. Lots of entanglement measures have been proposed, such as entanglement of formation, geometric measure of entanglement, concurrence, and  convex-roof extended negativity.

Consider a finite dimensional bipartite system, one is subsystem $A$ and the other one is subsystem $B$. The entanglement of formation (EOF) is the first entanglement measure built by the convex roof construction \cite{EOFde1,EOFde2}. For a pure state $|\psi\rangle$, it is defined by $E_F(|\psi\rangle)=S(\varrho_A)$, where $S(\varrho)=-\mathrm{Tr}(\varrho\log_2\varrho)$ stands for the von Neumann entropy and $\varrho_A=\tr_B(|\psi\rangle\langle\psi|)$ is the reduced density matrix of subsystem $A$. For a mixed state $\varrho$, the EOF is defined by the convex roof, $E_F(\varrho)=\inf_{\{p_i,|\psi_i\rangle\}}\sum_i p_i E_F(|\psi_i\rangle)$ for all possible ensemble realizations $\varrho=\sum_i p_i |\psi_i\rangle\langle\psi_i|$, where $p_i\geq0$ and $\sum_i p_i=1$. The infimum represents the minimal possible average entanglement over all pure state decompositions of $\varrho$.
The geometric measure of entanglement (GME) is another kind of convex-roof entanglement measures \cite{GME1,GME2}. For an arbitrary bipartite pure state $|\psi\rangle=U_A\otimes U_B \sum_i\sqrt{\mu_i}|ii\rangle$ with $\sqrt{\mu_i}$ being its Schmidt coefficients, the GME is defined by $E_G(|\psi\rangle)=1-\max\{\mu_i\}$. Similarly, the GME is extended to mixed states by the convex roof.
The concurrence was firstly introduced for two-qubit states by Ref. \cite{EOFde1}, and based on it Wootters and co-workers derived computable formulae for concurrence and EOF in the two-qubit case \cite{2qubit1,2qubit2}.
After that, Refs. \cite{concurrence3,concurrence4,concurrence5} extended it to bipartite higher-dimensional systems, i.e., $C(|\psi\rangle)=\sqrt{2(1-\tr\varrho_A^2)}$ for pure states, and its convex roof for mixed states.
The convex roof extended negativity (CREN) is generalized from the negativity, which is strongly related with the partial transpose \cite{ppt}. For a bipartite state $\varrho$, its negativity reads $N(\varrho)=\|\varrho^{T_B}\|-1$ \cite{negativity} (for simplicity we ignore the coefficient $1/2$ ), where $\|\cdot\|$ stands for the trace norm and $T_B$ is partial transpose with respect to subsystem $B$. The positive negativity is a necessary and sufficient condition of entanglement for pure states, $2\times 2$ and $2\times 3$ mixed states, but only a sufficient condition for higher-dimensional mixed states \cite{ppt}. To overcome this drawback, Lee \textit{et al.} proposed the CREN \cite{CREN}. For a pure state $|\psi\rangle$, CREN is defined by the negativity $\mathcal{N}(|\psi\rangle)=N(|\psi\rangle)=\||\psi\rangle\langle\psi|^{T_B}\|-1$. For mixed states, CREN is defined by the convex roof. Last but not least, the G-concurrence can be defined by the determinant of the reduced density matrix: $C_G(|\psi\rangle)=m(\det \varrho_A)^{1/m}$ for an $m\otimes n$ ($m\leq n$) pure state $|\psi\rangle$ \cite{GME1,Gour,Fan}, and for mixed states, the G-concurrence is defined by the convex roof as well \cite{Uhlmann}.

Although many entanglement measures have been proposed \cite{EOFde1,EOFde2,GME1,GME2,2qubit1,2qubit2,concurrence3,concurrence4,concurrence5,ppt,negativity,CREN,Gour,Fan,Uhlmann}, there are only a few explicit expressions of these measures for two-qubit states and some special kinds of higher-dimensional mixed states \cite{2qubit2,eof1,GME2,iso3,werner,CREN}. Furthermore, for a general state it is proved that computing many entanglement measures including the entanglement of formation is NP-hard \cite{NP,yichen}, which implies that we could only derive bounds on (rather than compute exact values of) these entanglement measures. Thus, in order to evaluate entanglement measures, lower and upper bounds of entanglement measures for general higher-dimensional states and multipartite states have been proposed \cite{mintert04,bounds1,bounds2,bounds3,bounds4,bounds5,bounds6,bounds7,bounds8,bounds9,bounds10,bounds11,bounds12,bounds13,bounds14,ob2,bounds15,bounds16}. Besides, the lower bounds of entanglement measures can server as a valuable tool for optimal control \cite{cui}.
However, if the proposed bound cannot be directly measured in experiments, quantum state tomography has to be performed which leads to rapidly growing experimental resources as system size increases. Therefore, experimentally observable lower and upper bounds of entanglement measures attract much interest recently \cite{mintert,guhne,lower,Huang,zhang,ob1,ob3,ob4,ob5,ob6}.

In this paper, we shall propose observable lower bounds for EOF, GME, concurrence, CREN, and G-concurrence in finite-dimensional bipartite systems. These lower bounds can be obtained from the expectation value of a single observable. Based on our results, we will present several examples using real experimental measurement data. Furthermore, the relations between entanglement measures will be studied.

\section{Entanglement of formation}
For simplicity, we use the denotation $\mathrm{co}(g)$. Here $\mathrm{co}(g)$ denotes the convex hull of the function $g$, which is the largest convex function that is bounded above by the given function $g$. The denotation has been used to get explicit expressions and bounds for the EOF \cite{eof1,eof2,eof3,svl2,eof4,eof5,werner}.

\textit{Theorem 1.---} For any $m\otimes n$ ($m\leq n$) quantum state $\varrho$, its entanglement of formation $E_F(\varrho)$ satisfies
\begin{eqnarray}\label{EOF}
E_F(\varrho)\geq\mathrm{co}[R(\Lambda)],
\end{eqnarray}
where $\Lambda=\max\{\langle\phi|\varrho|\phi\rangle/(s_1 m),1/m\}$, $R(\Lambda)=H_2[\gamma(\Lambda)]+[1-\gamma(\Lambda)]\log_2(m-1)$, $\gamma(\Lambda)=[\sqrt{\Lambda}+\sqrt{(m-1)(1-\Lambda)}]^2/m$,
with $H_2(x)=-x\log_2 x-(1-x)\log_2(1-x)$ being the standard binary entropy function, and $|\phi\rangle=V_A\otimes V_B \sum_{i=1}^{m}\sqrt{s_i}|ii\rangle$ being an arbitrary pure entangled state in $m\otimes n$ system (where $\{\sqrt{s_i}\}$ are its Schmidt coefficients in decreasing order). The convex hull of $R(\Lambda)$ is
\begin{eqnarray}
 \mathrm{co}[R(\Lambda)]&=&\left\{
\begin{array}{ll}
H_2[\gamma(\Lambda)]+[1-\gamma(\Lambda)]\log_2(m-1), \\
\Lambda \in \big[\frac{1}{m},\frac{4(m-1)}{m^2}\big]; \\[2mm]
\frac{m\log_2(m-1)}{m-2}(\Lambda-1)+\log_2 m, \\
\Lambda \in \big[\frac{4(m-1)}{m^2},1\big].
\end{array}%
\right.
\label{lowerbound}
\end{eqnarray}

\textit{Proof.--} We first find the minimal admissible $H(\vec{\mu})\equiv-\sum_{i=1}^m \mu_i\log_2\mu_i=-\mathrm{Tr}(\varrho\log_2\varrho)=S(\varrho)$ for a given $\lambda=(\sum_{i=1}^m\sqrt{\mu_i})^2/m$, where $\mu_i$ are eigenvalues of $\varrho$ and $\vec{\mu}$ is the Schmidt vector $\{\mu_1,\mu_2,\cdots,\mu_m\}$. Consider the following function,
\begin{eqnarray}
R(\lambda)&=&\min_{\vec{\mu}}\Bigg\{H(\vec{\mu})\bigg|\lambda=\frac{1}{m}\bigg(\sum_{i=1}^m\sqrt{\mu_i}\bigg)^2\Bigg\},\label{R2}
\end{eqnarray}
As shown in Ref. \cite{eof1}, the minimal $H(\vec{\mu})$ versus $\lambda$ corresponds to $\vec{\mu}$ in the form $\{t,(1-t)/(m-1),\cdots,(1-t)/(m-1)\}$ for $t\in[1/m,1]$. Therefore, one can get $R(\lambda)=H_2[\gamma(\lambda)]+[1-\gamma(\lambda)]\log_2(m-1)$ with $\gamma(\lambda)=[\sqrt{\lambda}+\sqrt{(m-1)(1-\lambda)}]^2/m$.

Suppose that we have already found an optimal decomposition $\sum_j p_j|\psi_j\rangle\langle\psi_j|$ for $\varrho$ to achieve the infimum of $E_F(\varrho)$, then $E_F(\varrho)=\sum_j p_j E_F(|\psi_j\rangle)$ by definition. Since $\mathrm{co}[R(\lambda)]$ is a monotonously increasing convex function and satisfies $\mathrm{co}[R(\lambda)]\leq R(\lambda)\leq H(\vec{\mu})$ for a given $\lambda$, one thus has
\begin{eqnarray}
E_F(\varrho)&=&\sum_j p_j E_F(|\psi_j\rangle)=\sum_j p_j H(\vec{\mu}^j)\nonumber\\
&\geq&\sum_j p_j \mathrm{co}[R(\lambda^j)]\geq \mathrm{co}[R(\sum_j p_j\lambda^j)]\geq\mathrm{co}[R(\Lambda)],\nonumber
\end{eqnarray}
where $|\psi_j\rangle=U_A\otimes U_B \sum_{i=1}^m \sqrt{\mu_i^j}|ii\rangle$ with $\{\sqrt{\mu_i^j}\}$ being its Schmidt coefficients in decreasing order,
and we have used
\begin{eqnarray}
\lambda^j&=&\frac{(\sum_{i=1}^m\sqrt{\mu_i^j})^2}{m}\geq\frac{(\sum_{i=1}^m\sqrt{s_i\mu_i^j})^2}{s_1 m}\nonumber\\
&\geq&\max_{U_1,U_2}\frac{\langle\phi|U_1\otimes U_2|\psi_j\rangle\langle\psi_j|U_1^\dag\otimes U_2^\dag|\phi\rangle}{s_1 m},
\end{eqnarray}
where the second inequality holds since the theorem shown in Ref. \cite{horn}, and the detailed proof has been given in the appendix. Therefore,
\begin{eqnarray}
\sum_j p_j\lambda^j&\geq& \max_{U_1,U_2}\frac{\langle\phi|U_1\otimes U_2\sum_j p_j|\psi_j\rangle\langle\psi_j|U_1^\dag\otimes U_2^\dag|\phi\rangle}{s_1 m}\nonumber\\
&\geq& \frac{\langle\phi|\varrho|\phi\rangle}{s_1 m}.
\end{eqnarray}
Together with $\lambda^j=(\sum_{i=1}^m\sqrt{\mu_i^j})^2/m\geq1/m$, one can get $\sum_j p_j\lambda^j\geq\max\{\langle\phi|\varrho|\phi\rangle/(s_1 m),1/m\}=\Lambda$. Since $\mathrm{co}[R(\lambda)]$ is a monotonously increasing function and $\sum_j p_j\lambda^j\geq\Lambda$, one has $\mathrm{co}[R(\sum_j p_j\lambda^j)]\geq\mathrm{co}[R(\Lambda)]$.

As introduced above, $\mathrm{co}(g)$ is the largest convex function that is bounded above by the given function $g$. From the expression of $R(\Lambda)$, the explicit expression of $\mathrm{co}[R(\Lambda)]$ is as Eq. (\ref{lowerbound}) \cite{eof1,eof2,eof3}. Actually, the same function has been gotten for the EOF of isotropic states \cite{eof1} and the lower bound of EOF based on partial transpose and realignment criteria in Refs. \cite{eof2,eof3}.   \hfill  $\blacksquare$


\section{Geometric measure of entanglement}
Similar to entanglement of formation, we can also find an observable lower bound for the GME.

\textit{Theorem 2.---} For any $m\otimes n$ ($m\leq n$) quantum state $\varrho$, its geometric measure of entanglement $E_G(\varrho)$ satisfies
\begin{equation}\label{GME}
    E_G(\varrho)\geq \mathrm{co}[Q(\Lambda)],
\end{equation}
where $\Lambda=\max\{\langle\phi|\varrho|\phi\rangle/(s_1 m),1/m\}$, $Q(\Lambda)=1-\gamma(\Lambda)$
with $\gamma(\Lambda)=[\sqrt{\Lambda}+\sqrt{(m-1)(1-\Lambda)}]^2/m$, and
\begin{equation}\label{}
  \mathrm{co}[Q(\Lambda)]=Q(\Lambda).
\end{equation}

\textit{Proof.---} We first find the minimal admissible $G(\vec{\mu})\equiv1-\mu_{max}$ for a given $\lambda=(\sum_{i=1}^m\sqrt{\mu_i})^2/m$, where $\mu_{max}=\max\{\mu_i\}$. Consider the following function,
\begin{eqnarray}
Q(\lambda)=\min_{\vec{\mu}}\Bigg\{G(\vec{\mu})\bigg|\lambda=\frac{1}{m}\bigg(\sum_{i=1}^m\sqrt{\mu_i}\bigg)^2\Bigg\}.\label{RU2}
\end{eqnarray}
As shown in the appendix, the minimal $G(\vec{\mu})$ versus $\lambda$ corresponds to $\vec{\mu}$ in the form $\{t,(1-t)/(m-1),\cdots,(1-t)/(m-1)\}$ for $t\in[1/m,1]$. Therefore, $Q(\lambda)=1-\gamma(\lambda)$ holds,
with $\gamma(\lambda)=[\sqrt{\lambda}+\sqrt{(m-1)(1-\lambda)}]^2/m$.

Similar to the proof of Theorem 1, suppose that we have already found an optimal decomposition $\sum_j p_j|\psi_j\rangle\langle\psi_j|$ for $\varrho$ to achieve the infimum of $E_G(\varrho)$, then $E_G(\varrho)=\sum_j p_j E_G(|\psi_j\rangle)$ by definition. Since $\mathrm{co}[Q(\lambda)]$ is a monotonously increasing convex function and satisfies $\mathrm{co}[Q(\lambda)]\leq Q(\lambda)\leq G(\vec{\mu})$ for a given $\lambda$, one thus has
\begin{eqnarray}
E_G(\varrho)&=&\sum_j p_j E_G(|\psi_j\rangle)=\sum_j p_j G(\vec{\mu}^j)\nonumber\\
&\geq&\sum_j p_j \mathrm{co}[Q(\lambda^j)] \geq\mathrm{co}[Q(\sum_j p_j\lambda^j)]\geq\mathrm{co}[Q(\Lambda)],\nonumber
\end{eqnarray}
where, similar to the proof of Theorem 1, we have used $\sum_j p_j\lambda^j\geq\Lambda$.

From the definition of $\mathrm{co}$, one can see $\mathrm{co}[Q(\Lambda)]=Q(\Lambda)$, since $Q(\Lambda)$ is a convex function.  The same function has been used for the GME of isotropic states \cite{GME2}. \hfill  $\blacksquare$

\section{Concurrence} In the following, we shall also present an observable lower bound for the concurrence.


\textit{Theorem 3.---} For any $m\otimes n$ ($m\leq n$) quantum state $\varrho$, its concurrence $C(\varrho)$ satisfies
\begin{equation}\label{concurrence}
    C(\varrho)\geq \mathrm{co}[P(\Lambda)],
\end{equation}
where  $P(\Lambda)=\sqrt{2[1-\gamma(\Lambda)][m\gamma(\Lambda)+m-2]/(m-1)}$,
with $\gamma(\Lambda)=[\sqrt{\Lambda}+\sqrt{(m-1)(1-\Lambda)}]^2/m$, $\Lambda=\max\{\langle\phi|\varrho|\phi\rangle/(s_1 m),1/m\}$, and
\begin{equation}\label{coP}
  \mathrm{co}[P(\Lambda)]=\sqrt{\frac{2m}{m-1}}(\Lambda-\frac{1}{m}).
\end{equation}

\textit{Proof.---} Similarly, we first find the minimal admissible $L(\vec{\mu})\equiv\sqrt{2(1-\sum_{i=1}^m\mu_i^2)}$ for a given $\lambda=(\sum_{i=1}^m\sqrt{\mu_i})^2/m$. Consider the following function,
\begin{eqnarray}
P(\lambda)&=&\min_{\vec{\mu}}\Bigg\{L(\vec{\mu})\bigg|\lambda=\frac{1}{m}\bigg(\sum_{i=1}^m\sqrt{\mu_i}\bigg)^2\Bigg\}.\label{C2}
\end{eqnarray}
As shown in the appendix, the minimal $L(\vec{\mu})$ versus $\lambda$ corresponds to $\vec{\mu}$ still in the form $\{t,(1-t)/(m-1),\cdots,(1-t)/(m-1)\}$ for $t\in[1/m,1]$. Therefore, $P(\lambda)=\sqrt{2[1-\gamma(\lambda)][m\gamma(\lambda)+m-2]/(m-1)}$ holds,
with $\gamma(\lambda)=[\sqrt{\lambda}+\sqrt{(m-1)(1-\lambda)}]^2/m$.

Similar to the proofs of Theorem 1 and 2, suppose that we have already found an optimal decomposition $\sum_j p_j|\psi_j\rangle\langle\psi_j|$ for $\varrho$ to achieve the infimum of $C(\varrho)$, then $C(\varrho)=\sum_j p_j C(|\psi_j\rangle)$ by definition. Since $\mathrm{co}[P(\lambda)]$ is a monotonously increasing convex function and satisfies $\mathrm{co}[P(\lambda)]\leq L(\vec{\mu})$ for a given $\lambda$, one thus has
\begin{eqnarray}
C(\varrho)&=&\sum_j p_j C(|\psi_j\rangle)=\sum_j p_j L(\vec{\mu}^j)\nonumber\\
&\geq&\sum_j p_j \mathrm{co}[P(\lambda^j)] \geq\mathrm{co}[P(\sum_j p_j\lambda^j)]\geq\mathrm{co}[P(\Lambda)],\nonumber
\end{eqnarray}
where $\sum_j p_j\lambda^j\geq\Lambda$ has been used again.

From the definition of co, one can get Eq. (\ref{coP}),
since $P(\Lambda)$ is a monotonously increasing concave function as shown in the appendix. The same function has been gotten for the concurrence of isotropic states \cite{iso3}, the lower bound of concurrence based on partial transpose and realignment criteria in Ref. \cite{kai}, and the special case of Eq. (\ref{concurrence}) with $|\phi\rangle=1/\sqrt{m}\sum_{i=1}^{m}|ii\rangle$ being the maximally entangled state \cite{fef1,fef2}.  \hfill  $\blacksquare$

\section{Convex-roof extended negativity} Similarly, an observable lower bound of CREN has been presented as follows.

\textit{Theorem 4.---} For any $m\otimes n$ ($m\leq n$) quantum state $\varrho$, its convex-roof extended negativity $\mathcal{N}(\varrho)$ satisfies
\begin{equation}\label{CREN}
    \mathcal{N}(\varrho)\geq m\Lambda-1,
\end{equation}
where $\Lambda=\max\{\langle\phi|\varrho|\phi\rangle/(s_1 m),1/m\}$.

\textit{Proof.---}
Similar to the proofs of the theorems above, suppose that we have already found an optimal decomposition $\sum_j p_j|\psi_j\rangle\langle\psi_j|$ for $\varrho$ to achieve the infimum of $\mathcal{N}(\varrho)$, then $\mathcal{N}(\varrho)=\sum_j p_j \mathcal{N}(|\psi_j\rangle)$ by definition. It is worth noticing that, for an arbitrary pure state $|\psi_j\rangle=U_A\otimes U_B \sum_i \sqrt{\mu_i^j}|ii\rangle$ with $\sqrt{\mu_i^j}$ being its Schmidt coefficients in decreasing order, we have $\mathcal{N}(|\psi_j\rangle)=(\sum_{i=1}^m\sqrt{\mu_i^j})^2-1$.
Thus,
\begin{eqnarray}
\mathcal{N}(\varrho)&=&\sum_j p_j \mathcal{N}(|\psi_j\rangle)=\sum_j p_j\Big(\big(\sum_{i=1}^m\sqrt{\mu_i^j}\big)^2-1\Big)\nonumber\\
&=&m\sum_j p_j\lambda^j-1\geq m\Lambda-1,\nonumber
\end{eqnarray}
where $\sum_j p_j\lambda^j\geq\Lambda$ has been proved in Theorem 1. Actually, a similar function has been gotten for the CREN of isotropic states \cite{CREN}. \hfill  $\blacksquare$

\section{G-concurrence} Last but not least, one can present an observable lower bound of G-concurrence as follows.

\textit{Theorem 5.---} For any $m\otimes n$ ($m\leq n$) quantum state $\varrho$, its G-concurrence $C_G(\varrho)$ satisfies
\begin{equation}\label{g-concurrence}
    C_G(\varrho)\geq \mathrm{co}[K(\Lambda)],
\end{equation}
where  $K(\Lambda)=m[\gamma(\Lambda) \beta(\Lambda)^{m-1}]^{1/m}$,
with $\gamma(\Lambda)=[\sqrt{\Lambda}-\sqrt{(m-1)(1-\Lambda)}]^2/m$, $\beta(\Lambda)=[\sqrt{\Lambda}+\sqrt{(1-\Lambda)/(m-1)}]^2/m$, $\Lambda=\max\{\langle\phi|\varrho|\phi\rangle/(s_1 m),1/m\}$, and
\begin{equation}\label{coK}
  \mathrm{co}[K(\Lambda)]=\max\{1-m(1-\Lambda),0\}.
\end{equation}

\textit{Proof.---} Similarly, we first find the minimal admissible $S(\vec{\mu})\equiv m(\prod_{i=1}^m \mu_i)^{1/m}$ for a given $\lambda=(\sum_{i=1}^m\sqrt{\mu_i})^2/m$. Consider the following function,
\begin{eqnarray}
K(\lambda)&=&\min_{\vec{\mu}}\Bigg\{S(\vec{\mu})\bigg|\lambda=\frac{1}{m}\bigg(\sum_{i=1}^m\sqrt{\mu_i}\bigg)^2\Bigg\}.\label{C2}
\end{eqnarray}
As shown in Ref. \cite{201605} and the appendix, the minimal $S(\vec{\mu})$ versus $\lambda$ corresponds to $\vec{\mu}$ in the form $\{t,(1-t)/(m-1),\cdots,(1-t)/(m-1)\}$ for $t\in[0,1/m]$. Therefore,  $K(\lambda)=m[\gamma(\lambda) \beta(\lambda)^{m-1}]^{1/m}$ holds,
with $\gamma(\lambda)=[\sqrt{\lambda}-\sqrt{(m-1)(1-\lambda)}]^2/m$ and $\beta(\lambda)=[\sqrt{\lambda}+\sqrt{(1-\lambda)/(m-1)}]^2/m$.

Similar to the proofs of above theorems, suppose that we have already found an optimal decomposition $\sum_j p_j|\psi_j\rangle\langle\psi_j|$ for $\varrho$ to achieve the infimum of $C_G(\varrho)$, then $C_G(\varrho)=\sum_j p_j C_G(|\psi_j\rangle)$ by definition. Since $\mathrm{co}[K(\lambda)]$ is a monotonously increasing convex function and satisfies $\mathrm{co}[K(\lambda)]\leq S(\vec{\mu})$ for a given $\lambda$, one thus has
\begin{eqnarray}
C_G(\varrho)&=&\sum_j p_j C_G(|\psi_j\rangle)=\sum_j p_j S(\vec{\mu}^j)\nonumber\\
&\geq&\sum_j p_j \mathrm{co}[K(\lambda^j)] \geq\mathrm{co}[K(\sum_j p_j\lambda^j)]\geq\mathrm{co}[K(\Lambda)],\nonumber
\end{eqnarray}
where $\sum_j p_j\lambda^j\geq\Lambda$ has been used again.

From the definition of co, one can get Eq. (\ref{coK}),
since $K(\lambda)$ is a monotonously increasing concave function in $[(m-1)/m,1]$ as shown in Ref. \cite{201605}. The same function has been gotten for the G-concurrence of axisymmetric states \cite{201605}, and the lower bound of G-concurrence for arbitrary states which is the special case of Eq. (\ref{g-concurrence}) with $|\phi\rangle=1/\sqrt{m}\sum_{i=1}^{m}|ii\rangle$ being the maximally entangled state \cite{201605}.  \hfill  $\blacksquare$

\textit{Remark 1.---} It is worth noticing that when we choose a special case for $|\phi\rangle$, i.e. $|\phi\rangle=|\psi^+\rangle$ where $|\psi^+\rangle=1/\sqrt{m}\sum_{i=1}^{m}|ii\rangle$ is the maximally entangled state, then $s_1$ becomes to $1/m$ and $\Lambda=\max\{\langle\psi^+|\varrho|\psi^+\rangle,1/m\}$. For this special case, our results shown in all the above theorems can be proved in a different manner: for an arbitrary $m\otimes m$ state $\varrho$, one can project it onto the isotropic states by the twirling operation, i.e. $\varrho^{iso}=\int \mathrm{d}U(U\otimes U^*)\varrho(U\otimes U^*)^\dag$, which is a local operations and classical communication (LOCC) operation and therefore cannot increase entanglement. Moreover, we have $\langle\psi^+|\varrho|\psi^+\rangle=\langle\psi^+|\varrho^{iso}|\psi^+\rangle$, since it is invariant under the twirling operation. The entanglement measures of the isotropic states can be expressed as a function of $\langle\psi^+|\varrho^{iso}|\psi^+\rangle$ \cite{eof1,GME2,iso3,CREN}. Thus, one can get lower bounds of entanglement measures for $\varrho$ from the entanglement measures of the isotropic states, this idea to get a lower bound has been known since the earliest paper \cite{EOFde1}. However, this alternative proof for $|\phi\rangle=|\psi^+\rangle$ is not valid for a general $|\phi\rangle$.

\textit{Remark 2.---} From Theorem 1 to Theorem 5, all the lower bounds proposed above are the functions of $\Lambda$, which only depends on the expectation value of a single observable $|\phi\rangle\langle\phi|$. Therefore, it will be much easier to evaluate than tomography in experiments.


\section{Examples and experimental measurements} The first example is a real experimental state shown in Ref. \cite{real}. Tonolini \textit{et al.} experimentally realized a high-dimensional two-photon entangled state, with dimension of each photon being equal to $d=17$. They reconstructed the density matrix $\varrho_{exp_1}$ of the experimental state, and found the fidelity with the maximally entangled pure state being $\tr\sqrt{\sqrt{\varrho_{exp_1}}|\psi^+\rangle\langle\psi^+|\sqrt{\varrho_{exp_1}}}=0.831$. Therefore, our parameter for this experimental state should be $\Lambda_1=\max\{\langle\psi^+|\varrho_{exp_1}|\psi^+\rangle,1/17\}\doteq0.69$. Using Eqs. (\ref{EOF}), (\ref{GME}), (\ref{concurrence}), and (\ref{CREN}), one can arrive at
\begin{eqnarray}
&&E_F(\varrho_{exp_1})\geq2.68, \ \ \ \ E_G(\varrho_{exp_1})\geq0.45, \nonumber\\
&&C(\varrho_{exp_1})\geq0.92,  \ \ \ \ \ \mathcal{N}(\varrho_{exp_1})\geq10.73, \nonumber
\end{eqnarray}
for this real experimental state.

The second example is also from a real experiment \cite{real2}. In Ref. \cite{real2}, the authors experimentally realized a special three-photon pure state: $|\psi_{s}\rangle=\frac{\sqrt{3}}{2}|000\rangle+\frac{\sqrt{3}}{4}|110\rangle+\frac{1}{4}|111\rangle$ and a four-photon Dicke state with two exciations $|D_{4}^{2}\rangle=\frac{1}{\sqrt{6}}(|0011\rangle+|0101\rangle+|0110\rangle+|1001\rangle+|1010\rangle+|1100\rangle)$. The square of fidelities for the special state and the Dicke state are measured to be $\langle\psi_{s}|\varrho_{exp_2}|\psi_{s}\rangle=0.9821$ and $\langle D_4^2|\varrho_{exp_3}|D_4^2\rangle=0.9780$, respectively. For simplicity, we only consider the entanglement under $A|BC$ bipartition for the three-photon special state and $AB|CD$ bipartition for the four-photon Dicke state. Other bipartition entanglement can also be calculated analytically based on our theorems. Therefore, our parameters for these experimental states should be $\Lambda_2=\max\{\langle\psi_{s}|\varrho_{exp_2}|\psi_{s}\rangle/(s_1 m),1/m\}\doteq0.6547$ with $m=2$ and $s_1=3/4$, and $\Lambda_3=\max\{\langle D_4^2|\varrho_{exp_3}|D_4^2\rangle/(s_1 m),1/m\}\doteq0.3667$ with $m=4$ and $s_1=2/3$. Using Eqs. (\ref{EOF}), (\ref{GME}), (\ref{concurrence}), and (\ref{CREN}), one can arrive at
\begin{eqnarray}
&&E_F(\varrho_{exp_2})\geq0.1661, \ \ \ \ E_G(\varrho_{exp_2})\geq0.0245, \nonumber\\
&&C(\varrho_{exp_2})\geq0.3094,  \ \ \ \ \ \mathcal{N}(\varrho_{exp_2})\geq0.3094, \nonumber\\
&&C_G(\varrho_{exp_2})\geq0.3094, \nonumber\\
&&E_F(\varrho_{exp_3})\geq0.1437, \ \ \ \ E_G(\varrho_{exp_3})\geq0.0160, \nonumber\\
&&C(\varrho_{exp_3})\geq0.1905,  \ \ \ \ \ \mathcal{N}(\varrho_{exp_3})\geq0.4668, \nonumber
\end{eqnarray}
for these real experimental states.

In the first example shown above, Tonolini \textit{et al.} used compressive sensing to reduce the number of measurements, and 2506 measurements are needed for state reconstruction. Actually, the whole state reconstruction is not necessary for our lower bounds. We only need to perform $d^2$ local measurements (or one global measurement) to obtain the expectation value of $|\phi\rangle\langle\phi|$ for $d\times d$ dimensional states. If the global measurement $|\phi\rangle\langle\phi|$ is not easy to be performed in experiments, one can use local measurements instead of it. We recall a complete set of local orthogonal observables (LOOs) introduced in Ref. \cite{LOO}, i.e., $\{G_k\}=\{|l\rangle\langle l|,(|m\rangle\langle n|+|n\rangle \langle m|)/\sqrt{2},(|m\rangle\langle n|-|n\rangle \langle m|)/(i\sqrt{2})\}$, where $1\leq l\leq d$ and $1\leq m<n\leq d$. The LOOs satisfy the orthogonal condition $\tr (G_k G_{k'})=\delta_{kk'}$, and the complete condition $\hat{A}=\sum_k \tr(G_k \hat{A})G_k$ for an arbitrary observable $\hat{A}$. Thus,
$ |\phi\rangle\langle\phi|=\sum_{k=1}^{d^2}\langle\phi|V_A G_k V_A^\dag\otimes V_B G_k^T V_B^\dag|\phi\rangle V_A G_k V_A^\dag\otimes V_B G_k^T V_B^\dag$,
where $G_k^T$ means transpose of $G_k$ which is another complete set of LOOs, and $\{\langle \phi|V_A G_k V_A^\dag\otimes V_B G_k^T V_B^\dag|\phi\rangle\}=\{s_l,\sqrt{s_n s_m},\sqrt{s_n s_m}\}$. Therefore, only $d^2$ local measurements are required which is much less than the number of measurements $d^4$ required by tomography. For the first example $d=17$, we only need $289$ local measurements, much less than 2506 measurements needed by the compressive sensing. When $|\phi\rangle=|\psi^+\rangle$, the expression of $|\psi^+\rangle\langle\psi^+|$ has been widely known from Ref. \cite{nielsen}. Moreover, Ref. \cite{namiki} proposed methods to get a lower bound of the maximally entangled fraction with a rather smaller number of local measurements.

\section{Relations between entanglement measures} There have been comparative studies of entanglement measures. Horodecki \textit{et al.} introduced axiomatic approach for entanglement measures \cite{Horodecki}. Eltschka \textit{et al.} proposed inequalities between the concurrence and CREN for any $m\otimes n$ ($m\leq n$) quantum state $\varrho$ \cite{fef2}, $\mathcal{N}(\varrho)\geq C(\varrho)\geq\sqrt{2/[m(m-1)]}\mathcal{N}(\varrho)$.
We shall study the relations between concurrence, geometric measure of entanglement and CREN.

\textit{Theorem 6.---} For any $m\otimes n$ ($m\leq n$) quantum state $\varrho$, its concurrence $C(\varrho)$, geometric measure of entanglement $E_G(\varrho)$, and convex-roof extended negativity $\mathcal{N}(\varrho)$ satisfy
\begin{eqnarray}
&&\mathcal{N}(\varrho)\leq m\gamma\big(1-E_G(\varrho)\big)-1,\label{N}\\
&&E_G(\varrho)\geq1-\gamma\bigg(\frac{\mathcal{N}(\varrho)+1}{m}\bigg),\label{EG}\\
&&c(\varrho)^2+\big(1-e_G(\varrho)\big)^2\leq1,\label{tradeoff}
\end{eqnarray}
where $\gamma(x)=[\sqrt{x}+\sqrt{(m-1)(1-x)}]^2/m$ with $x\in[1/m,1]$ and the integer $m\geq2$, $c(\varrho)=\sqrt{m/[2(m-1)]}C(\varrho)$, $e_G(\varrho)=mE_G(\varrho)/(m-1)$.

\textit{Proof.---}
Let $\{p_j,|\psi_j\rangle\}$ be the optimal ensemble for $E_G(\varrho)$ and $\sqrt{\mu_i^j}$ be the Schmidt coefficients in decreasing order for $|\psi_j\rangle$. Thus,
\begin{eqnarray}
\mathcal{N}(\varrho)&\leq&\sum_j p_j\mathcal{N}(|\psi_j\rangle)=\sum_j p_j\bigg(\sum_{i=1}^m\sqrt{\mu_i^j}\bigg)^2-1\nonumber\\
&\leq&\sum_j p_j\bigg(\sqrt{\mu_1^j}+\sqrt{(m-1)(1-\mu_1^j)}\bigg)^2-1\nonumber\\
&\leq&m\gamma\big(\sum_j p_j\mu_1^j\big)-1=m\gamma\big(1-E_G(\varrho)\big)-1,\nonumber
\end{eqnarray}
since $\gamma(x)$ is a concave function and we have used the fact that $\sum_{i}\sqrt{\mu_i}=\sqrt{\mu_1}+\sum_{i>1}\sqrt{\mu_i}\leq\sqrt{\mu_1}+\sqrt{(m-1)(\sum_{i>1}\mu_i)}=\sqrt{\mu_1}+\sqrt{(m-1)(1-\mu_1)}$.

In order to prove Eq. (\ref{EG}), the properties of $\gamma(x)$ are needed: $\gamma(x)$ is a decreasing function, i.e., $\gamma(x)\geq\gamma(y)$ if $x\leq y$, and $\gamma(\gamma(x))=x$. Therefore, from Eq. (\ref{N}) we have $\gamma[\mathcal{N}(\varrho)/m+1/m]\geq1-E_G(\varrho)$,
which is equivalent to Eq. (\ref{EG}).

It is worth noticing that $\sum_i\mu_i^2=\mu_1^2+\sum_{i>1}\mu_i^2\geq\mu_1^2+(\sum_{i>1}\mu_i)^2/(m-1)=1-2\bar{\mu}_1+m\bar{\mu}_1^2/(m-1)$ where $\bar{\mu}_1=1-\mu_1$.
Thus, $m(1-\sum_i\mu_i^2)/(m-1)\leq2m\bar{\mu}_1/(m-1)-m^2\bar{\mu}_1^2/(m-1)^2=1-[1-m\bar{\mu}_1/(m-1)]^2$ holds.
Together with $\bar{\mu}_1\leq(m-1)/m$, we have $1-m\bar{\mu}_1/(m-1)\leq\sqrt{1-m(1-\sum_i\mu_i^2)/(m-1)}$.
Therefore,
\begin{eqnarray}
e_G(\varrho)
&\geq&\sum_j p_j\bigg(1-\sqrt{1-\frac{m}{m-1}(1-\sum_i{\mu_i^j}^2)}\bigg)\nonumber\\
&\geq&1-\sqrt{1-\frac{m}{m-1}\sum_j p_j(1-\sum_i{\mu_i^j}^2)}\nonumber\\
&\geq&1-\sqrt{1-c(\varrho)^2}, \nonumber
\end{eqnarray}
holds, and since $c(\varrho),e_G(\varrho)\in[0,1]$ we get Eq. (\ref{tradeoff}). \hfill  $\blacksquare$

\section{Discussion and conclusion} Actually, the method we used to get lower bounds of EOF, GME, concurrence, CREN and G-concurrence can be generalized to arbitrary bipartite convex-roof entanglement measures. Suppose that the entanglement measure on an $m\otimes n$ ($m\leq n$) pure state $|\psi\rangle$ is $E(|\psi\rangle)=f(\vec{\mu})$, where $\vec{\mu}$ is the Schmidt vector. We first get the minimal admissible $f(\vec{\mu})$ for a given $\lambda=(\sum_{i=1}^m\sqrt{\mu_i})^2/m$, i.e., $F(\lambda)=\min_{\vec{\mu}}\{f(\vec{\mu})|\lambda=(\sum_{i=1}^m\sqrt{\mu_i})^2/m\}$. Thus, for a general $m\otimes n$ ($m\leq n$) state $\varrho$, the lower bound of this entanglement measure $E(\varrho)$ is given by $E(\varrho)\geq\mathrm{co}[F(\Lambda)]$ with $\Lambda=\max\{\langle\phi|\varrho|\phi\rangle/(s_1 m),1/m\}$, if the final function $\mathrm{co}[F(\lambda)]$ is a monotonously increasing convex function with respect to $\lambda$.

Our paper actually provided lower bounds on a variety of bipartite entanglement measures from a well-known entanglement
witness \cite{exp1}:
\begin{eqnarray}
W=s_1\idol-|\phi\rangle\langle\phi|,
\end{eqnarray}
where $|\phi\rangle=V_A\otimes V_B \sum_{i=1}^{m}\sqrt{s_i}|ii\rangle$ is an arbitrary pure entangled state in $m\otimes n$ system (with $\{\sqrt{s_i}\}$ being its Schmidt coefficients in decreasing order). This entanglement witness has been measured in many experiments, such as Refs. \cite{exp1,exp2,exp3,exp4,exp5,exp6,exp7,exp8}. Using this entanglement witness (where $|\phi\rangle$ can be an arbitrary pure entangled state rather than only a maximally entangled state), one can obtain lower bounds on a variety of bipartite entanglement measures based on our method.


In conclusion, we present observable lower bounds for several entanglement measures defined by convex roof, which include EOF, GME, concurrence, CREN, and G-concurrence. The lower bounds estimate these entanglement measures for arbitrary finite-dimensional bipartite states. Moreover, these lower bounds can be easily obtained from the expectation value of a single observable. Based on our results, we present some examples using real experimental measurement data. In principle, our method can be used for arbitrary finite-dimensional bipartite convex-roof entanglement measures. Last but not least, the relations between entanglement measures are studied.

\section*{ACKNOWLEDGMENTS}
This work is funded by the Singapore Ministry of Education (partly through the Academic Research Fund Tier 3 MOE2012-T3-1-009), the National Research Foundation, Singapore (Grant No. WBS: R-710-000-008-271), the financial support from RGC of Hong Kong(Grant No. 538213), the National Natural Science Foundation of China (Grants No. 11504253 and No. 11575155), and the startup funding from Soochow University (Grant No. Q410800215).

\setcounter{equation}{0}
\renewcommand{\theequation}{S\arabic{equation}}

\section*{APPENDIX}

Here we provide some details of calculations to get the expressions of $R(\lambda)$, $\mathrm{co}[R(\lambda)]$, $Q(\lambda)$, $\mathrm{co}[Q(\lambda)]$, $P(\lambda)$, $\mathrm{co}[P(\lambda)]$, $K(\lambda)$ and $\mathrm{co}[K(\lambda)]$. The main idea is to calculate lower bounds of entanglement measures as a function of $\lambda$ for pure states, and then to extend the bounds to mixed states by convex hull. The details of proving the inequality $\sum_j p_j \lambda^j\geq\Lambda$ have also been presented.

\subsection{Calculation of $R(\lambda)$ and $\mathrm{co}[R(\lambda)]$}
In the following, we shall seek the minimal $H(\vec{\mu})=-\sum_{i=1}^m \mu_i\log_2\mu_i$ for a given $\lambda=(\sum_{i=1}^m\sqrt{\mu_i})^2/m$. We use $R(\lambda)$ to denote the minimal $H(\vec{\mu})$ for a given $\lambda$, i.e.,
\begin{eqnarray}
R(\lambda)&=&\min_{\vec{\mu}}\Bigg\{H(\vec{\mu})\bigg|\lambda=\frac{1}{m}\bigg(\sum_{i=1}^m\sqrt{\mu_i}\bigg)^2\Bigg\},\label{RR2}
\end{eqnarray}
As shown in Ref. \cite{eof1}, the minimal $H(\vec{\mu})$ versus $\lambda$ corresponds to $\vec{\mu}$ in the form
\begin{equation}\label{}
    \vec{\mu}=\bigg\{t,\frac{1-t}{m-1},\cdots,\frac{1-t}{m-1}\bigg\}  \ \ \ \mathrm{for} \ t\in\bigg[\frac{1}{m},1\bigg],
\end{equation}
with $m-1$ copies of $(1-t)/(m-1)$ and one copy of $t$.
Therefore,  the minimal $H(\vec{\mu})$ and corresponding $\lambda$ are
\begin{eqnarray}
H(t)&=&-t\log_2 t-(1-t)\log_2 \frac{1-t}{m-1},\label{H}\\
\lambda(t)&=&\frac{1}{m}\Big(\sqrt{t}+\sqrt{(1-t)(m-1)}\Big)^2.
\end{eqnarray}
In order to show the minimal $H(\vec{\mu})$ versus $\lambda$, we need the inverse function of $\lambda(t)$. After some algebra, one can arrive at
\begin{eqnarray}
t(\lambda)&=&\frac{1}{m}\Big(\sqrt{\lambda}+\sqrt{(m-1)(1-\lambda)}\Big)^2,\label{t}
\end{eqnarray}
with $\lambda\in[1/m,1]$. Substituting Eq. (\ref{t}) into Eq. (\ref{H}), we can get the expression for $R(\lambda)$.

\begin{figure}
\begin{center}
\includegraphics[scale=0.6]{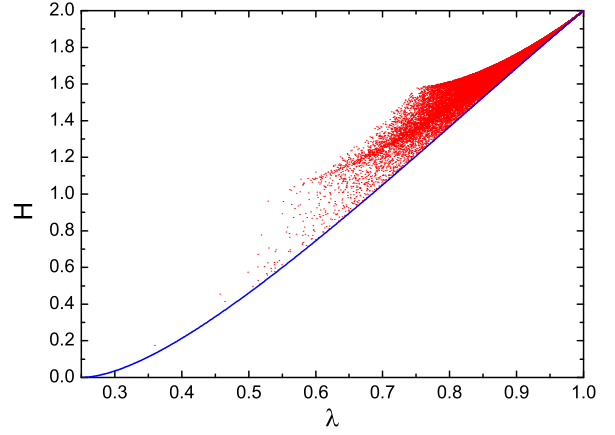}
\caption{(Color online) $H(\vec{\mu})$ versus $\lambda$. 50000 red dots represent randomly generated states with $m=4$. The lower boundary is a smooth blue curve, which corresponds to $R(\lambda)$.}\label{1}
\end{center}
\end{figure}

We can simulate the lower boundary of the region in $H(\vec{\mu})$ versus $\lambda$ plane. 50000 dots for randomly generated states with $m=4$ are displayed in Fig. \ref{1}. The lower boundary corresponds to $R(\lambda)$. It is defined that $\mathrm{co}(g)$ is the largest convex function that is bounded above by the given function $g$. From the expression of $R(\Lambda)$,  the explicit expression of $\mathrm{co}[R(\Lambda)]$ is obtained in Refs. \cite{eof1,eof2,eof3}, which is Eq. (2) shown in the main text.

\subsection{Calculation of $Q(\lambda)$ and $\mathrm{co}[Q(\lambda)]$}
We first seek the minimal $G(\vec{\mu})\equiv1-\mu_{max}$ for a given $\lambda=(\sum_{i=1}^m\sqrt{\mu_i})^2/m$, where $\mu_{max}=\max\{\mu_i\}$. We use $Q(\lambda)$ to denote the minimal $G(\vec{\mu})$ for a given $\lambda$, i.e.,
\begin{eqnarray}
Q(\lambda)=\min_{\vec{\mu}}\Bigg\{G(\vec{\mu})\bigg|\lambda=\frac{1}{m}\bigg(\sum_{i=1}^m\sqrt{\mu_i}\bigg)^2\Bigg\}.\label{}
\end{eqnarray}
It is interesting that, similar to $H(\vec{\mu})$, the minimal $G(\vec{\mu})$ versus $\lambda$ corresponds to $\vec{\mu}$ in the form
\begin{equation}\label{}
    \vec{\mu}=\bigg\{t,\frac{1-t}{m-1},\cdots,\frac{1-t}{m-1}\bigg\}  \ \ \ \mathrm{for} \ t\in\bigg[\frac{1}{m},1\bigg],
\end{equation}
with $m-1$ copies of $(1-t)/(m-1)$ and one copy of $t$.
Therefore,  the minimal $G(\vec{\mu})$ and corresponding $\lambda$ are
\begin{eqnarray}
G(t)&=&1-t,\label{G}\\
\lambda(t)&=&\frac{1}{m}\Big(\sqrt{t}+\sqrt{(1-t)(m-1)}\Big)^2,
\end{eqnarray}
where $\mu_{max}=t$, since $t\geq(1-t)/(m-1)$ with $t\in[1/m,1]$.
In order to show the minimal $G(\vec{\mu})$ versus $\lambda$, we need the inverse function of $\lambda(t)$. After some algebra, one can see
that
\begin{eqnarray}
t(\lambda)&=&\frac{1}{m}\Big(\sqrt{\lambda}+\sqrt{(m-1)(1-\lambda)}\Big)^2,\label{tt}
\end{eqnarray}
with $\lambda\in[1/m,1]$. Substituting Eq. (\ref{tt}) into Eq. (\ref{G}), we can get the expression for $Q(\lambda)$, i.e.,
\begin{eqnarray}
Q(\lambda)=1-\frac{1}{m}\Big(\sqrt{\lambda}+\sqrt{(m-1)(1-\lambda)}\Big)^2.\label{SQ}
\end{eqnarray}

From Eq. (\ref{SQ}), we can find that
\begin{eqnarray}
\frac{\mathrm{d}Q(\lambda)}{\mathrm{d}\lambda}=\frac{\bigg(\sqrt{\frac{m-1}{1-\lambda}}-\frac{1}{\sqrt{\lambda}}\bigg)\Big(\sqrt{\lambda}+\sqrt{(m-1)(1-\lambda)}\Big)}{m}
\geq0,\nonumber
\end{eqnarray}
when $\lambda\in[1/m,1]$. Thus, $Q(\lambda)$ is a monotonously increasing function. Moreover, from the definition of $\mathrm{co}()$, one can see that $\mathrm{co}[Q(\lambda)]=Q(\lambda)$, since $Q(\lambda)$ is a convex function. In order to prove this, we only need to show that $\mathrm{d}^2Q(\lambda)/\mathrm{d}\lambda^2\geq0$. From Eq. (\ref{SQ}), one can get
\begin{eqnarray}
\frac{\mathrm{d}^2Q(\lambda)}{\mathrm{d}\lambda^2}=\frac{\sqrt{(m-1)(1-\lambda)}}{2m(1-\lambda)^2\lambda^{3/2}}\geq0,
\end{eqnarray}
since $m$ is an integer ($m\geq2$) and $\lambda\in[1/m,1]$. Therefore, $\mathrm{co}[Q(\lambda)]=Q(\lambda)$, which is a monotonously increasing convex function.

\begin{figure}
\begin{center}
\includegraphics[scale=0.6]{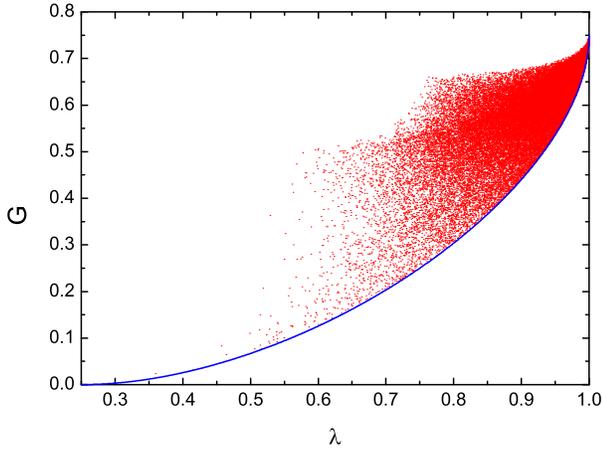}
\caption{(Color online) $G(\vec{\mu})$ versus $\lambda$. 50000 red dots represent randomly generated states with $m=4$. The lower boundary is a smooth blue curve, which corresponds to both $Q(\lambda)$ and $\mathrm{co}[Q(\lambda)]$, since $\mathrm{co}[Q(\lambda)]=Q(\lambda)$.}\label{2}
\end{center}
\end{figure}

We can simulate the lower boundary of the region in $G(\vec{\mu})$ versus $\lambda$ plane. In Fig. \ref{2}, 50000 dots for randomly generated states with $m=4$ are displayed. The lower boundary corresponds to $Q(\lambda)$, which coincides with $\mathrm{co}[Q(\lambda)]$.

\subsection{Calculation of $P(\lambda)$ and $\mathrm{co}[P(\lambda)]$}
We first seek the minimal $L(\vec{\mu})\equiv\sqrt{2(1-\sum_{i=1}^m\mu_i^2)}$ for a given $\lambda=(\sum_{i=1}^m\sqrt{\mu_i})^2/m$. We use $P(\lambda)$ to denote the minimal $L(\vec{\mu})$ for a given $\lambda$, i.e.,
\begin{eqnarray}
P(\lambda)=\min_{\vec{\mu}}\Bigg\{L(\vec{\mu})\bigg|\lambda=\frac{1}{m}\bigg(\sum_{i=1}^m\sqrt{\mu_i}\bigg)^2\Bigg\}.\label{}
\end{eqnarray}
It is interesting that, similar to $H(\vec{\mu})$ and $G(\vec{\mu})$, the minimal $L(\vec{\mu})$ versus $\lambda$ corresponds to $\vec{\mu}$ in the form
\begin{equation}\label{}
    \vec{\mu}=\bigg\{t,\frac{1-t}{m-1},\cdots,\frac{1-t}{m-1}\bigg\}  \ \ \ \mathrm{for} \ t\in\bigg[\frac{1}{m},1\bigg],
\end{equation}
with $m-1$ copies of $(1-t)/(m-1)$ and one copy of $t$.
Therefore, the minimal $L(\vec{\mu})$ and corresponding $\lambda$ are
\begin{eqnarray}
L(t)&=&\sqrt{\frac{2(1-t)(mt+m-2)}{m-1}},\label{L}\\
\lambda(t)&=&\frac{1}{m}\Big(\sqrt{t}+\sqrt{(1-t)(m-1)}\Big)^2.
\end{eqnarray}
In order to show the minimal $L(\vec{\mu})$ versus $\lambda$, we need the inverse function of $\lambda(t)$. After some algebra, one can arrive at
\begin{eqnarray}
t(\lambda)&=&\frac{1}{m}\Big(\sqrt{\lambda}+\sqrt{(m-1)(1-\lambda)}\Big)^2,\label{ttt}
\end{eqnarray}
with $\lambda\in[1/m,1]$. Substituting Eq. (\ref{ttt}) into Eq. (\ref{L}), we can get the expression for $P(\lambda)$, i.e.,
\begin{eqnarray}
P(\lambda)&=&\sqrt{\frac{2(1-t)(mt+m-2)}{m-1}},\label{SP}\\
t&=&\frac{1}{m}\Big(\sqrt{\lambda}+\sqrt{(m-1)(1-\lambda)}\Big)^2.\label{SP2}
\end{eqnarray}

From Eqs. (\ref{SP}) and (\ref{SP2}), we can find that
\begin{eqnarray}
\frac{\mathrm{d}P(\lambda)}{\mathrm{d}\lambda}=\frac{\mathrm{d}P(t)}{\mathrm{d}t}\frac{\mathrm{d}t(\lambda)}{\mathrm{d}\lambda},
\end{eqnarray}
where
\begin{eqnarray}
\frac{\mathrm{d}P(t)}{\mathrm{d}t}&=&\frac{\sqrt{2}(1-mt)}{\sqrt{(m-1)(1-t)(m+mt-2)}}\leq0,\\
\frac{\mathrm{d}t(\lambda)}{\mathrm{d}\lambda}&=&\frac{1}{m}\Bigg(\frac{1}{\sqrt{\lambda}}+\frac{1-m}{\sqrt{(m-1)(1-\lambda)}}\Bigg)\nonumber\\
&&\times\Big(\sqrt{\lambda}+\sqrt{(m-1)(1-\lambda)}\Big)\leq0,
\end{eqnarray}
since $m\geq2$ and $t,\lambda\in[1/m,1]$. Therefore,
\begin{eqnarray}
\frac{\mathrm{d}P(\lambda)}{\mathrm{d}\lambda}\geq0,
\end{eqnarray}
which means $P(\lambda)$ is a monotonously increasing function.

\begin{figure}
\begin{center}
\includegraphics[scale=0.6]{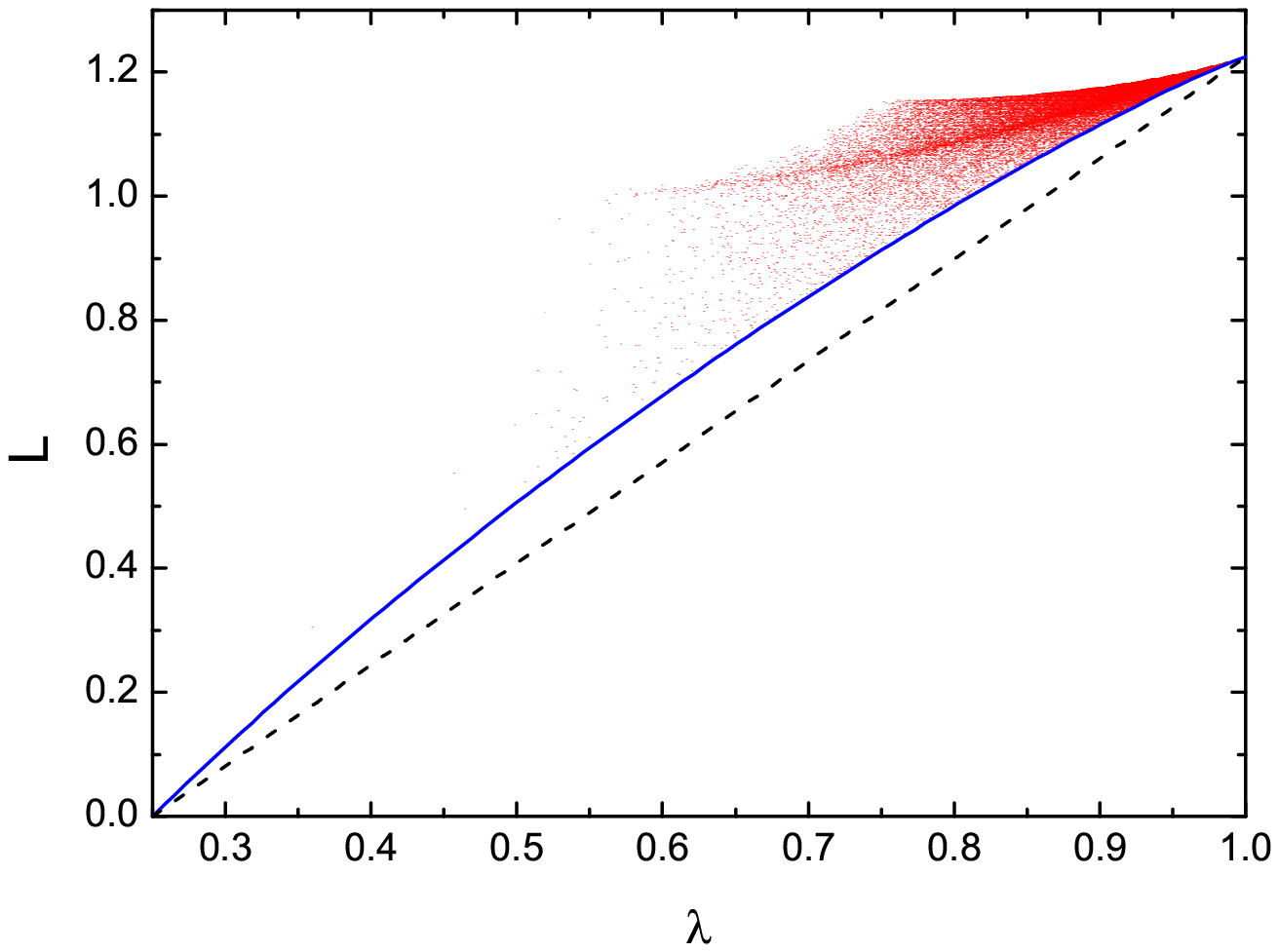}
\caption{(Color online) $L(\vec{\mu})$ versus $\lambda$. 50000 red dots represent randomly generated states with $m=4$. The lower boundary is a smooth blue curve which corresponds to $P(\lambda)$, and the dashed black line corresponds to $\mathrm{co}[P(\lambda)]$.}\label{3}
\end{center}
\end{figure}

Furthermore, in order to show $P(\lambda)$ is a concave function, we need to prove that $\mathrm{d}^2P(\lambda)/\mathrm{d}\lambda^2\leq0$. One can use
\begin{eqnarray}
\frac{\mathrm{d}^2P(\lambda)}{\mathrm{d}\lambda^2}=\frac{\mathrm{d}^2P(t)}{\mathrm{d}t^2}\Bigg(\frac{\mathrm{d}t(\lambda)}{\mathrm{d}\lambda}\Bigg)^2+\frac{\mathrm{d}P(t)}{\mathrm{d}t}\frac{\mathrm{d}^2t(\lambda)}{\mathrm{d}\lambda^2}
\end{eqnarray}
and
\begin{eqnarray}
\frac{\mathrm{d}^2P(t)}{\mathrm{d}t^2}&=&-\sqrt{2}\Big(\frac{m-1}{(1-t)(m+mt-2)}\Big)^{3/2},\\
\frac{\mathrm{d}^2t(\lambda)}{\mathrm{d}\lambda^2}&=&-\frac{\sqrt{(m-1)(1-\lambda)}}{2m(1-\lambda)^2\lambda^{3/2}}.
\end{eqnarray}
Thus,
\begin{eqnarray}
&&\frac{\mathrm{d}^2P(\lambda)}{\mathrm{d}\lambda^2}\nonumber\\
&=&\frac{b^4(m-2)[(m-1)a-b]^2 f}{\sqrt{2}m^2(m-1)^{5/2}(1-\lambda)^2\lambda^{3/2}(1-t)^{3/2}(m+mt-2)^{3/2}},\nonumber
\end{eqnarray}
where
\begin{eqnarray}
  a&:=&\sqrt{\lambda},\\
  b&:=&\sqrt{(m-1)(1-x)},\\
  x&:=&\frac{a}{b},\\
  f&:=&-3(m-1)x^2+2(3-m)x+\frac{2m-3}{m-1},
\end{eqnarray}
with $x\in[1/(m-1),+\infty)$. It is easy to see that
\begin{equation}\label{}
    f\leq0,
\end{equation}
when $x\in[1/(m-1),+\infty)$. Thus,
\begin{equation}\label{}
    \frac{\mathrm{d}^2P(\lambda)}{\mathrm{d}\lambda^2}\leq0,
\end{equation}
when $m\geq2$, $t,\lambda\in[1/m,1]$, and $x\in[1/(m-1),+\infty)$. Therefore, the convex hull of $P(\lambda)$ will be
\begin{equation}
  \mathrm{co}[P(\lambda)]=\sqrt{\frac{2m}{m-1}}(\lambda-\frac{1}{m}),
\end{equation}
which is a straight line from $(1/m,0)$ to $(1,\sqrt{2(m-1)/m})$.

We can simulate the lower boundary of the region in $L(\vec{\mu})$ versus $\lambda$ plane. 50000 dots for randomly generated states with $m=4$ are displayed in Fig. \ref{3}. The lower boundary corresponds to $P(\lambda)$, and the dashed black line corresponds to $\mathrm{co}[P(\lambda)]$.

\subsection{Calculation of $K(\lambda)$ and $\mathrm{co}[K(\lambda)]$}
We first seek the minimal $S(\vec{\mu})\equiv m(\prod_{i=1}^m \mu_i)^{1/m}$ for a given $\lambda=(\sum_{i=1}^m\sqrt{\mu_i})^2/m$. We use $K(\lambda)$ to denote the minimal $S(\vec{\mu})$ for a given $\lambda$, i.e.,
\begin{eqnarray}
K(\lambda)=\min_{\vec{\mu}}\Bigg\{S(\vec{\mu})\bigg|\lambda=\frac{1}{m}\bigg(\sum_{i=1}^m\sqrt{\mu_i}\bigg)^2\Bigg\}.\label{}
\end{eqnarray}
As shown in Ref. \cite{201605}, the minimal $S(\vec{\mu})$ versus $\lambda$ corresponds to $\vec{\mu}$ in the form
\begin{equation}\label{}
    \vec{\mu}=\bigg\{t,\frac{1-t}{m-1},\cdots,\frac{1-t}{m-1}\bigg\}  \ \ \ \mathrm{for} \ t\in\bigg[0,\frac{1}{m}\bigg],
\end{equation}
with $m-1$ copies of $(1-t)/(m-1)$ and one copy of $t$.
Therefore, the minimal $S(\vec{\mu})$ and corresponding $\lambda$ are
\begin{eqnarray}
S(t)&=&m\bigg(t\frac{(1-t)^{m-1}}{(m-1)^{m-1}}\bigg)^{\frac{1}{m}},\label{S}\\
\lambda(t)&=&\frac{1}{m}\Big(\sqrt{t}+\sqrt{(1-t)(m-1)}\Big)^2.
\end{eqnarray}
In order to show the minimal $S(\vec{\mu})$ versus $\lambda$, we need the inverse function of $\lambda(t)$. After some algebra, one can arrive at
\begin{eqnarray}
t(\lambda)&=&\frac{1}{m}\Big(\sqrt{\lambda}-\sqrt{(m-1)(1-\lambda)}\Big)^2,\label{tttt}
\end{eqnarray}
with $\lambda\in[(m-1)/m,1]$. Substituting Eq. (\ref{tttt}) into Eq. (\ref{S}), we can get the expression for $K(\lambda)$, i.e.,
\begin{eqnarray}
K(\lambda)&=&m[t (1-t)^{m-1}]^{1/m},\\
t&=&\frac{1}{m}\Big(\sqrt{\lambda}-\sqrt{(m-1)(1-\lambda)}\Big)^2.
\end{eqnarray}

From Ref. \cite{201605}, one can see that $K(\lambda)$ is a monotonously increasing concave function in $[(m-1)/m,1]$. Therefore, the convex hull of $K(\lambda)$ will be
\begin{equation}
  \mathrm{co}[K(\lambda)]=\max\{1-m(1-\lambda),0\}.
\end{equation}

\begin{figure}
\begin{center}
\includegraphics[scale=0.9]{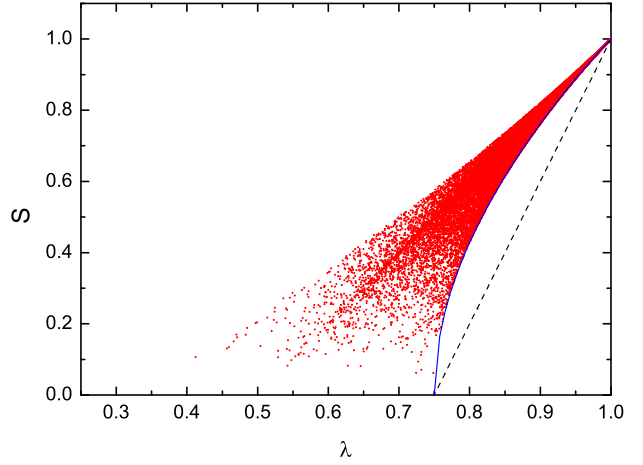}
\caption{(Color online) $S(\vec{\mu})$ versus $\lambda$. 50000 red dots represent randomly generated states with $m=4$. The lower boundary is a smooth blue curve which corresponds to $K(\lambda)$, and the dashed black line corresponds to $\mathrm{co}[K(\lambda)]$.}\label{4}
\end{center}
\end{figure}

We can simulate the lower boundary of the region in $S(\vec{\mu})$ versus $\lambda$ plane. 50000 dots for randomly generated states with $m=4$ are displayed in Fig. \ref{4}. The lower boundary corresponds to $K(\lambda)$, and the dashed black line corresponds to $\mathrm{co}[K(\lambda)]$.

\subsection{Proof of the inequality $\sum_j p_j \lambda^j\geq\Lambda$}
Here we give the details of proving $\sum_j p_j \lambda^j\geq\Lambda$. For any $m\otimes n$ ($m\leq n$) quantum state $\varrho$, suppose that we have found an optimal decomposition $\sum_j p_j|\psi_j\rangle\langle\psi_j|$ for $\varrho$ to achieve the infimum of $E(\varrho)$, where $E(\varrho)$ is one kind of entanglement measure defined by the convex roof. For each $|\psi_j\rangle$, we have the expression
\begin{eqnarray}
|\psi_j\rangle=U_A\otimes U_B \sum_{i=1}^m \sqrt{\mu_i^j}|ii\rangle=U_A\otimes U_B \sum_{i=1}^m\sum_{k=1}^n \Sigma_{ik}|ik\rangle,\nonumber
\end{eqnarray}
with $\{\sqrt{\mu_i^j}\}$ being its Schmidt coefficients in decreasing order, where $U_A$ ($U_B$) is an $m\times m$ ($n\times n$) unitary matrix and $\Sigma$ is an $m\times n$ matrix defined by $\Sigma_{ik}=\sqrt{\mu_i^j}\delta_{ik}$. Similarly, for an arbitrary pure entangled given state $|\phi\rangle$ in $m\otimes n$ system, we have the expression
\begin{eqnarray}
|\phi\rangle=V_A\otimes V_B \sum_{i=1}^{m}\sqrt{s_i}|ii\rangle=V_A\otimes V_B \sum_{i=1}^m\sum_{k=1}^n S_{ik}|ik\rangle,
\end{eqnarray}
where $\{\sqrt{s_i}\}$ are its Schmidt coefficients in decreasing order, $V_A$ ($V_B$) is an $m\times m$ ($n\times n$) unitary matrix, and $S$ is an $m\times n$ matrix defined by $S_{ik}=\sqrt{s_i}\delta_{ik}$. Therefore, the maximum under all possible $U_1$ (which is an $m\times m$ unitary matrix) and $U_2$ (which is an $n\times n$ unitary matrix) is
\begin{widetext}
\begin{eqnarray}
\max_{U_1,U_2}\frac{\langle\phi|U_1\otimes U_2|\psi_j\rangle\langle\psi_j|U_1^\dag\otimes U_2^\dag|\phi\rangle}{s_1 m}
&=&\max_{U_1,U_2}\frac{|\sum_{i,i'=1}^m\sum_{k,k'=1}^n\Sigma_{ik}S_{i'k'}^* \langle i'|V_A^\dag U_1 U_A|i\rangle\langle k'|V_B^\dag U_2 U_B|k\rangle|^2}{s_1 m}\nonumber\\
&=&\max_{U_1,U_2}\frac{|\tr(\mathcal{U}_1\Sigma\mathcal{U}_2^T S^\dag)|^2}{s_1 m}\nonumber\\
&\leq&\max_{U_1,U_2}\frac{|\sum_{i=1}^m \sigma_i(\mathcal{U}_1\Sigma\mathcal{U}_2^T S^\dag)|^2}{s_1 m}\nonumber\\
&\leq&\frac{|\sum_{i=1}^m \sigma_{i}(\mathcal{U}_1\Sigma)\sigma_i(\mathcal{U}_2^T S^\dag)|^2}{s_1 m}\nonumber\\
&=&\frac{(\sum_{i=1}^m\sqrt{s_i\mu_i^j})^2}{s_1 m},
\end{eqnarray}
where $\mathcal{U}_1$ is an $m\times m$ unitary matrix defined by ${\mathcal{U}_1}_{i'i}=\langle i'|V_A^\dag U_1 U_A|i\rangle$, $\mathcal{U}_2$ is an $n\times n$ unitary matrix defined by ${\mathcal{U}_2}_{k'k}=\langle k'|V_B^\dag U_2 U_B|k\rangle$, and $\{\sigma_i(A)\}$ are singular values of $A$ in decreasing order. The first inequality holds since $|\tr A|\leq\sum_i \sigma_i(A)$ for arbitrary matrix $A$, and the second inequality holds since the following theorem \cite{horn}:

\textit{Let $A$ ($n\times p$ matrix) and $B$ ($p\times m$ matrix) be given, let $q=\min\{n,p,m\}$, and denote the ordered singular values of $A$, $B$, and $AB$ by $\sigma_1(A)\geq\cdots\geq\sigma_{\min\{n,p\}}(A)\geq0$,  $\sigma_1(B)\geq\cdots\geq\sigma_{\min\{p,m\}}(B)\geq0$, and $\sigma_1(AB)\geq\cdots\geq\sigma_{\min\{n,m\}}(AB)\geq0$. Then
\begin{eqnarray}
\sum_{i=1}^q\sigma_i(AB)\leq\sum_{i=1}^q\sigma_i(A)\sigma_i(B).
\end{eqnarray}}

Therefore,
\begin{eqnarray}
\lambda^j=\frac{(\sum_{i=1}^m\sqrt{\mu_i^j})^2}{m}\geq\frac{(\sum_{i=1}^m\sqrt{s_i\mu_i^j})^2}{s_1 m}\geq\max_{U_1,U_2}\frac{\langle\phi|U_1\otimes U_2|\psi_j\rangle\langle\psi_j|U_1^\dag\otimes U_2^\dag|\phi\rangle}{s_1 m},
\end{eqnarray}
where the first inequality holds since $s_1=\max\{s_i\}$. Thus,
\begin{eqnarray}
\sum_j p_j\lambda^j\geq \max_{U_1,U_2}\frac{\langle\phi|U_1\otimes U_2\sum_j p_j|\psi_j\rangle\langle\psi_j|U_1^\dag\otimes U_2^\dag|\phi\rangle}{s_1 m}
\geq \frac{\langle\phi|\varrho|\phi\rangle}{s_1 m}.
\end{eqnarray}
Together with $\lambda^j=(\sum_{i=1}^m\sqrt{\mu_i^j})^2/m\geq1/m$, one can obtain $\sum_j p_j\lambda^j\geq\max\{\langle\phi|\varrho|\phi\rangle/(s_1 m),1/m\}=\Lambda$.
\end{widetext}


\end{document}